\documentclass[conference]{IEEEtran}
\IEEEoverridecommandlockouts
\usepackage{cite}
\usepackage[numbers]{natbib}
\usepackage{amsmath,amssymb,amsfonts}
\usepackage{algorithmic}
\usepackage{graphicx}
\usepackage{textcomp}
\usepackage{xcolor}
\def\BibTeX{{\rm B\kern-.05em{\sc i\kern-.025em b}\kern-.08em
    T\kern-.1667em\lower.7ex\hbox{E}\kern-.125emX}}
\usepackage{booktabs}
\usepackage{multirow}
\usepackage{subcaption}
\usepackage{color}
\usepackage{colortbl}
\usepackage{adjustbox}

\usepackage{standalone}
\usepackage{pgfplots}
\usepackage{pgf-pie}
\usepackage{threeparttable}
\usepackage[listings,skins,breakable]{tcolorbox}
\pgfplotsset{compat=1.8}
\usepgfplotslibrary{statistics}
\usepackage{multirow} 
\usepackage{url}
\usepackage{listings}
\usepackage{csquotes}
\usepackage{amsmath}
\usepackage[normalem]{ulem}
\newtcolorbox{quotebox}{colback=steel!10,boxrule=0.4pt,colframe=black,fonttitle=\bfseries,top=2pt,bottom=2pt}

\newtcolorbox{expbox}{colback=red!10,boxrule=0.4pt,colframe=black,fonttitle=\bfseries,top=2pt,bottom=2pt}

\definecolor{steel}{rgb}{0, 0.2, 0.9} 

  \newcommand{\quart}[4]{\begin{adjustbox}{max width=.1\textwidth}\begin{picture}(100,5)
    {\color{black}\put(#1,5){\line(1,0){#2}}\color{magenta}\put(#3,5){\circle*{7}}\color{black}\put(#3,5){\circle{7}}}\end{picture}\end{adjustbox}}
    
          \newcommand{\quartexp}[4]{\begin{adjustbox}{max width=.1\textwidth}\begin{picture}(20,5)
    {\color{black}\put(#1,3){\line(1,0){#2}}\color{magenta}\put(#3,3){\circle*{4}}\color{black}\put(#3,3){\circle{4}}}\end{picture}\end{adjustbox}}

\newenvironment{code-example}
{
\vspace{0.15cm}
\noindent\begin{minipage}{\linewidth}
\begin{center}
\arrayrulecolor{black}
\color{black}
\begin{tabular}{|p{0.95\linewidth}|}
\hline%
\rowcolor{pink!20}%
}
{
\\\hline
\end{tabular}
\end{center}
\end{minipage}
\vspace{-0.2cm}
}

\newcommand{\approach}{\texttt{MHNurf}}
\newcommand{\bugtype}{non-functional }
\newcommand{\Bugtype}{Non-functional }
\newcommand{\BugType}{Non-Functional }

\begin{document}

\title{Multifaceted Hierarchical Report Identification for \BugType Bugs in Deep Learning Frameworks}

\author{\IEEEauthorblockN{Guoming Long}
\IEEEauthorblockA{\textit{Department of Computer Science} \\
\textit{Loughborough University}\\
Loughborough, United Kingdom \\
g.long@lboro.ac.uk}
\and
\IEEEauthorblockN{Tao Chen\IEEEauthorrefmark{1}}
\IEEEauthorblockA{\textit{Department of Computer Science} \\
\textit{Loughborough University}\\
Loughborough, United Kingdom \\
t.t.chen@lboro.ac.uk}
\and
\IEEEauthorblockN{Georgina Cosma}
\IEEEauthorblockA{\textit{Department of Computer Science} \\
\textit{Loughborough University}\\
Loughborough, United Kingdom \\
g.cosma@lboro.ac.uk}}

\maketitle
\begingroup\renewcommand\thefootnote{\IEEEauthorrefmark{1}}
\footnotetext{Corresponding author}
\endgroup
\begin{abstract}
\Bugtype bugs (e.g., performance- or accuracy-related bugs) in Deep Learning (DL) frameworks can lead to some of the most devastating consequences. Reporting those bugs on a repository such as GitHub is a standard route to fix them. Yet, given the growing number of new GitHub reports for DL frameworks, it is intrinsically difficult for developers to distinguish those that reveal \bugtype bugs among the others, and assign them to the right contributor for investigation in a timely manner. In this paper, we propose \approach~--- an end-to-end tool for automatically identifying \bugtype bug related reports in DL frameworks. The core of \approach~is a Multifaceted Hierarchical Attention Network (MHAN) that tackles three unaddressed challenges: (1) learning the semantic knowledge, but doing so by (2) considering the hierarchy (e.g., words/tokens in sentences/statements) and focusing on the important parts (i.e., words, tokens, sentences, and statements) of a GitHub report, while (3) independently extracting information from different types of features, i.e., \textit{content}, \textit{comment}, \textit{code}, \textit{command}, and \textit{label}. 

To evaluate \approach, we leverage 3,721 GitHub reports from five DL frameworks for conducting experiments. The results show that \approach~works the best with a combination of \textit{content}, \textit{comment}, and \textit{code}, which considerably outperforms the classic HAN where only the \textit{content} is used. \approach~also produces significantly more accurate results than nine other state-of-the-art classifiers with strong statistical significance, i.e., up to 71\% AUC improvement and has the best Scott-Knott rank on four frameworks while 2nd on the remaining one. To facilitate reproduction and promote future research, we have made our dataset, code, and detailed supplementary results publicly available at: \textcolor{blue}{\texttt{\url{https://github.com/ideas-labo/APSEC2022-MHNurf}}}.
\end{abstract}


\section{Introduction}
\label{sec:intro}

Deep learning (DL), which is a kind of machine intelligence algorithms that mimics the workings of the human brain in processing data~\cite{goodfellow2016deep}, has been gaining momentum in both academia and industry~\citep{DBLP:conf/nips/KrizhevskySH12,DBLP:journals/tse/ChenB17,DBLP:conf/issta/Liu0020,DBLP:conf/msr/GongC22,DBLP:conf/icse/Chen19b}. As such, several well-known DL frameworks (e.g., \texttt{TensorFlow}, \texttt{Keras}, and \texttt{PyTorch}) were created and maintained on GitHub, aiming to provide effective and readily available API for seamlessly adopting the DL algorithms into real-world problems. 


Despite the success of DL frameworks, they inevitably contain bugs, which, if left unfixed, would propagate issues to any applications that were built on top of them~\cite{DBLP:conf/kbse/GuoXLZLLS20}. Among other bugs, there exist \bugtype bugs that have no explicit symptoms of exceptions (such as a Not-a-Number error or the program crashes), i.e., they cannot be judged by using a precise oracle. For instance, common examples of \bugtype bugs are performance- or accuracy-related bugs (which is the focus of this work), since from the perspective of the DL frameworks, it is typically hard to understand how ``slow'' or how ``inaccurate'' the results are would be considered as a bug without thorough investigation, therefore they are more challenging to be analyzed. However, those \bugtype bugs tend to cause some of the most devastating outcomes and hence are of great concern~\cite{DBLP:conf/sigsoft/0001L21,DBLP:journals/tosem/ChenLBY18}. Indeed, according to the U.S. National Transportation Safety Board (NTSB), the recent accident of Uber's self-driving car was caused by a \bugtype bug of their DL framework, which classified the pedestrian as an unknown object with a slow reaction\footnote{\url{https://tinyurl.com/ykufbpey}.}.

To deal with bugs, it is a normal Software Engineering practice for DL frameworks to allow users to submit a report on repositories like GitHub, which would then be assigned to a contributor for formal investigation with an attempt to fix the bug, if any~\cite{DBLP:conf/fase/BadashianHS16}. Identifying whether a report is \bugtype bugs related (among other functional counterparts) is a labor-intensive process. This is because firstly, the number of new reports increases dramatically. For example, there are around 700 monthly new GitHub reports for \texttt{Tensorflow} in average\footnote{https://github.com/tensorflow/tensorflow/pulse/monthly}, including bugs related ones and those for other purposes, such as feature requests and help seeking. Secondly, GitHub reports can be lengthy, e.g., it could be up to 332 sentences per report on average~\cite{DBLP:conf/sigsoft/ManiCSD12}. Finally, given the vague nature of \bugtype bug, it is fundamentally difficult to understand if the related reports really reflect bugs. The above mean that, when assigning or prioritizing the GitHub reports, it can take a long time for developers to read and understand the bug reports, hence delaying the potential fixes to the destructive \bugtype bugs, especially when some of the key messages are deeply hidden inside. 

In light of the above, the problem we focus on in this paper is the following: given a GitHub report for DL framework, can we automatically learn and identify whether it is a \bugtype bug related report? Indeed, many existing classifiers on bug report identification can be directly applied. For example, those that identify a particular type of bug report~\cite{DBLP:journals/infsof/GomesTC21} (e.g., long-lived bugs); those that predict whether a bug report is bug-related~\cite{DBLP:journals/ese/HerboldTT20}; and those that classify reports based on labels~\cite{DBLP:journals/tr/UmerLI20,DBLP:conf/icmla/0003FWBL18,DBLP:conf/icsm/KallisSCP19}. However, in addition to the fact that these works do not target the level of DL frameworks, they have failed to handle some or all of the following challenges, which are important in the report identification:

\begin{itemize} 
    \item \textbf{Semantics matter:} Depending on the context, the same words or code tokens in the GitHub reports can have different meanings. Existing classifiers using statistical learning algorithms~\cite{DBLP:journals/infsof/GomesTC21} could fail to handle this polysemy.
    
     \item \textbf{Multiple types of features exist:} While most existing classifiers consider the content (\textit{title} and \textit{description}) of a GitHub report~\cite{DBLP:journals/infsof/GomesTC21,DBLP:journals/ese/HerboldTT20,DBLP:journals/tr/UmerLI20,DBLP:conf/icmla/0003FWBL18,DBLP:conf/icsm/KallisSCP19}, other types of features may also provide useful information, such as the accumulated comments made by the participants before a contributor is assigned to the report. Further, the mix of code and natural language in a report can pose additional challenge. 
     
     \item \textbf{Not all parts are equally relevant:} Given a lengthy GitHub report, not all of the words and sentences are important in identifying the \bugtype bug related reports. Yet, existing work has often ignored such a fact~\cite{DBLP:journals/tr/UmerLI20,DBLP:conf/icmla/0003FWBL18,DBLP:conf/icsm/KallisSCP19}. 
\end{itemize}

In this paper, we propose \textbf{\underline{M}}ultifaceted \textbf{\underline{H}}ierarchical \textbf{\underline{N}}on-functional B\textbf{\underline{u}}g Repo\textbf{\underline{r}}t Identi\textbf{\underline{f}}ication, dubbed \approach~--- an end to end tool for automatically identifying \bugtype bug related reports for DL frameworks. Its core component is a newly proposed Multifaceted Hierarchical Attention Network (MHAN) in this work, which extends the Hierarchical Attention Network (HAN)~\cite{DBLP:conf/naacl/YangYDHSH16}.


\textbf{Contributions:} To better identify \bugtype bug related reports for DL frameworks, our contributions include:

\begin{itemize}
    \item \approach~learns the semantic knowledge by considering the hierarchy and discriminating important and unimportant parts in GitHub reports. 
    \item The MHAN in \approach~considers multifacetedness in GitHub reports, i.e., it learns up to five types of feature (\textit{content}, \textit{comment}, \textit{code}, \textit{command}, and \textit{label}) independently.
    \item By using a dataset of 3,721 GitHub reports from five DL frameworks (i.e., \texttt{TensorFlow}, \texttt{PyTorch}, \texttt{Keras}, \texttt{MXNet}, and \texttt{Caffe}), the experiment results confirm that the \textit{title} and \textit{description}, which are fundamental parts of a GitHub report, needs to be considered together as part of the \textit{content} feature.
    \item We also found that the combination of \textit{content}, \textit{comment}, and \textit{code} give the best result in \approach. Notably, the multifacetedness in \approach~has also helped to considerably improve the prediction against the vanilla HAN in \approach.
    \item By comparing with nine state-of-the-art classifiers, we show that \approach~is significantly more accurate in general (up to 71\% AUC improvement). It is also ranked as the 1st for 4 out of 5 DL frameworks; 2nd for one, according to the Scott-Knott test~\cite{DBLP:journals/tse/MittasA13}. We also conduct a qualitative analysis on why \approach~can perform better.
\end{itemize}

\section{Problem Context and Challenges}
\label{sec:prob}

\subsection{Context}
DL frameworks hosted on GitHub allows participants and users to submit issue reports, whose purpose includes, but not limited to, bugs, pull request, feature request, and others. Among those, the GitHub reports that are bug-related are often of high importance to the community, especially the \bugtype bug related reports. Here, we distinguish two types of users/developers on GitHub:

\begin{itemize}
    \item \textbf{Contributors:} who are assigned to a report so that the formal investigation starts.
    \item \textbf{Participants:} who have not been assigned to a report, but are free to make comments on it.
\end{itemize}

Most commonly, after assignments, those GitHub reports need to be reviewed by the contributors who will pick the most important ones to investigate and fix when necessary.

\begin{figure}[t!]
  \centering
  \includegraphics[width=\columnwidth]{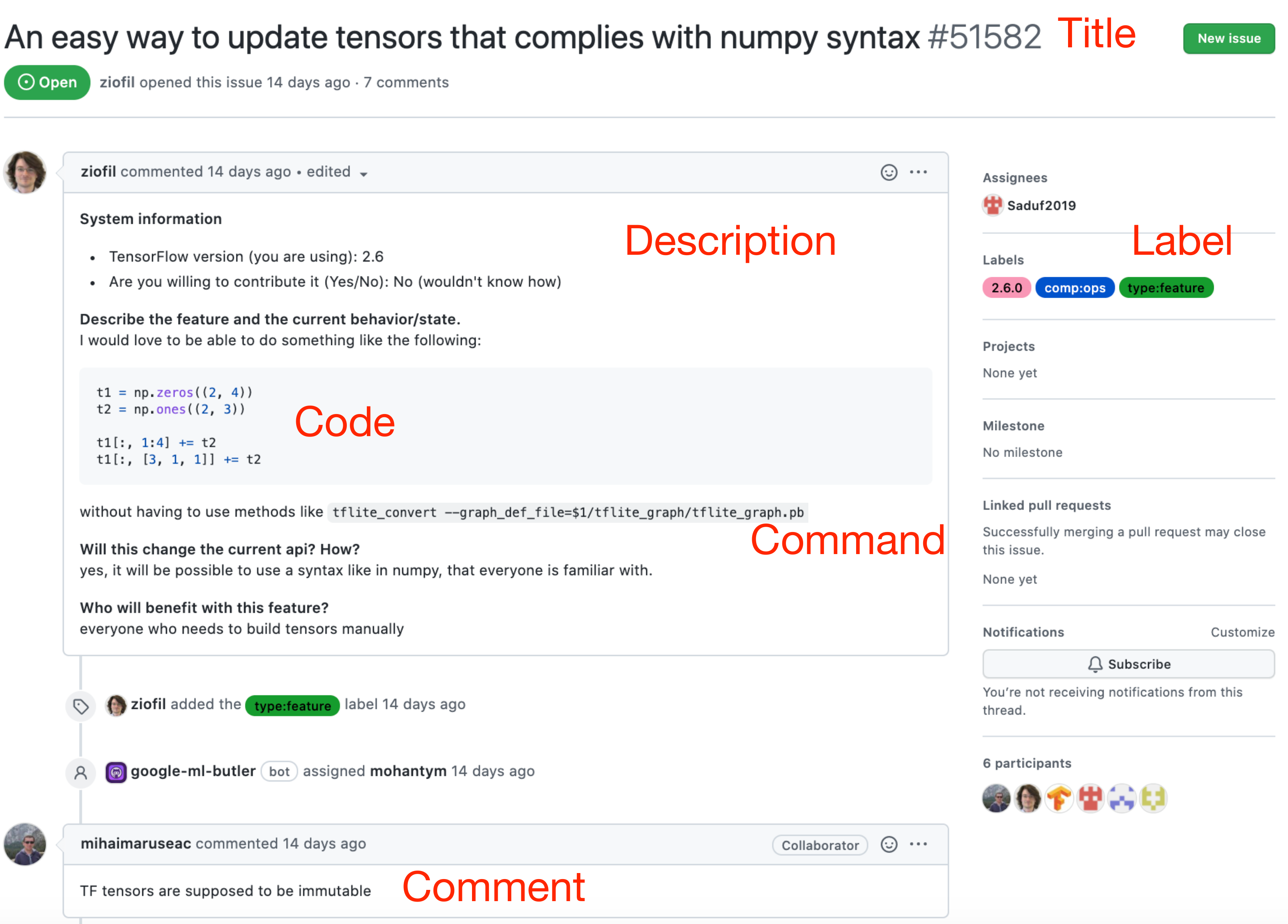}
  \caption{An example of GitHub report from \texttt{TensorFlow}.}
  \label{fig:exp}
\end{figure}

As shown in Figure~\ref{fig:exp}, apart from the normal content (\textit{title} and \textit{description}), a GitHub report may likely be commented on by different participants before being assigned to a contributor. Further, a label may also be added by the automatic bot or by a contributor based on his/her first impression. A formal investigation begins when the report is assigned to someone.

Our work here is to automatically identify those GitHub reports that are \bugtype bug related for a DL framework. This need not have to be done immediately when a report is submitted but also can be achieved in a short period of time after submission, as long as it is before the assignment. This can then provide useful information for the bot (or human) who assigns the GitHub reports in terms of which contributor to assign and how to prioritize them, hence saving more maintenance efforts and mitigating the consequence of truly devastating bugs. 

In this regard, we are dealing with a binary classification problem: a GitHub report can be either describing a \bugtype bug (e.g., performance and/or accuracy related) or not, which may be a functional bug report or reveal no bug at all (e.g., it is a feature request, API misuse, or merely a false alarm).

\subsection{Key Challenges}
\label{sec:chall}
Here we describe the key challenges identified about the GitHub reports for DL frameworks, which motivate our designs of \approach.


\textbf{\underline{Challenge 1: Semantics matter.}} GitHub reports for DL frameworks also contain strong semantic information, especially for \bugtype bugs. For example, in report \texttt{\#26736} for \texttt{TesnorFlow}, words like \texttt{\textcolor{blue}{`slow'}} is the key to indicate that it is a (performance) \bugtype bug related report. However, this does not mean that any report with a term \texttt{\textcolor{blue}{`slow'}} is \bugtype bug related, as the context and semantic information are also important. In another example --- report \texttt{\#63854} for \texttt{PyTorch} --- the \texttt{\textcolor{blue}{`slow'}} merely means a particular mode under which a logical bug was explored.

The above are common examples where the semantics matter for identifying \bugtype bug related reports for DL frameworks.


\textbf{\underline{Challenge 2: Multiple types of features exist.}} As shown in Figure~\ref{fig:exp}, a report for DL frameworks on GitHub consists of different types of features, which we summarize as follows:

\begin{itemize}
    \item \textbf{Content:} This is the fundamental feature for a report on GitHub, which includes both {title} and {description}. They are often the most important source of information in \bugtype bug related report identification.
    \item \textbf{Comment:} For GitHub reports, comments from the participants may be accumulated before a contributor is assigned.
    \item \textbf{Code:} Code for reproduction is also a common feature in the description of a GitHub report. In particular, here we refer to the problematic code, test code, the results returned, and the stack traces that are quoted in the code blocks as part of the description.
    \item \textbf{Command:} This is the code snippet that is not part of the code block but is mixed with the texts in the description.
    \item \textbf{Label:} Every GitHub report may come with one or more labels, such as \texttt{runtime} and \texttt{build}, which may be added by a human based on initial impression.
\end{itemize}

\textbf{\underline{Challenge 3: Not all parts are equally relevant.}} Another intuition we found is that not all parts in the GitHub report are equally important for identifying whether it is related to \bugtype bugs. For example, in Table~\ref{tb:word-exp}, only the highlighted sentence provides the most important information for one to understand whether the corresponding report is \bugtype bug related or not. The words \texttt{\textcolor{blue}{`speed'}} and \texttt{\textcolor{blue}{`slowed'}} imply more about a performance problem, hence it should be related to non-functional bugs.

\begin{table}[t!]
\caption{Texts from the content of \bugtype bug related report \texttt{\#33340} for \texttt{TensorFlow}. The \setlength{\fboxsep}{0.5pt}\colorbox{pink}{red} sentence gives strong meaning of a \bugtype bug report while the \setlength{\fboxsep}{0.5pt}\colorbox{steel!30}{blue} words \texttt{`speed'} and \texttt{`slowed'} contribute the most therein.}
\label{tb:word-exp}
\centering
\small
\begin{tabular}{p{8.1cm}}

\toprule
\setlength{\fboxsep}{1pt}\colorbox{pink}{The prediction \setlength{\fboxsep}{0.5pt}\colorbox{steel!30}{speed} is \setlength{\fboxsep}{0.5pt}\colorbox{steel!30}{slowed} down after \texttt{model.compile()}.} Predict function is used by users assuming that it will work fast because we use it all the time in production. It should not cause surprise to users.\\
\bottomrule
\end{tabular}
\end{table}


\begin{figure}[t!]
  \centering
  \includegraphics[width=\columnwidth]{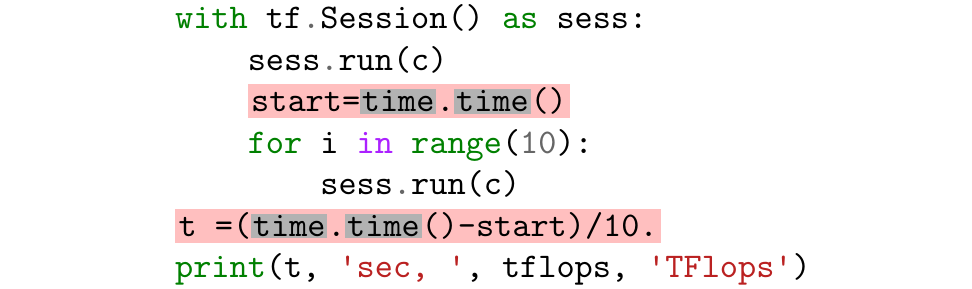}
  \caption{Code snippet extracted from \bugtype bug related report \texttt{\#27623} for \texttt{TensorFlow}. The \setlength{\fboxsep}{0.5pt}\colorbox{pink}{red} statements give more clues to indicate a \bugtype bug report while the \setlength{\fboxsep}{0.5pt}\colorbox{steel!30}{blue} token \texttt{`time'} contributes the most therein.}
  \label{fig:code-exp}
\end{figure}

Similarly, for code, as shown in Figure~\ref{fig:code-exp}, the statements \texttt{\textbf{\textcolor{blue}{start}}} \texttt{\textbf{\textcolor{blue}{=time.time()}}} and \texttt{\textbf{\textcolor{blue}{t =(time.time()-start)/10}}} indicate that it is important to record running time in the code snippet, which is a clear indication that performance is of great concern in the report. In particular, the token \texttt{\textbf{\textcolor{blue}{time}}} is what contributes more clues for one to confirm whether the report is \bugtype bug related --- a strong sign that this can be a \bugtype bug related report.

As a result, appropriate attention needs to be paid to different parts and their hierarchy in the report for identifying \bugtype bug-related ones in DL frameworks.

\section{Multifaceted Hierarchical Attention Network}
\label{sec:net}
In this section, we elaborate on the key component underpins \approach, namely Multifaceted Hierarchical Attention Network (MHAN), which modfiies the Hierarchical Attention Network (HAN)~\cite{DBLP:conf/naacl/YangYDHSH16}. The idea is to adopt multiple attention mechanisms to learn the semantic knowledge collaboratively (\textbf{Challenge 1}). This is achieved by independently extracting the rich semantic information in different feature types of the DL GitHub reports (hence handling the multifacetedness in \textbf{Challenge 2}) while paying hierarchical attention to the important words and sentences (or tokens and statements) within each (hence tackling the unequal importance and information hierarchy in \textbf{Challenge 3}). These independent pieces of information would then be concatenated to make a prediction.

\begin{figure}[t!]
  \centering
  \includegraphics[width=\columnwidth]{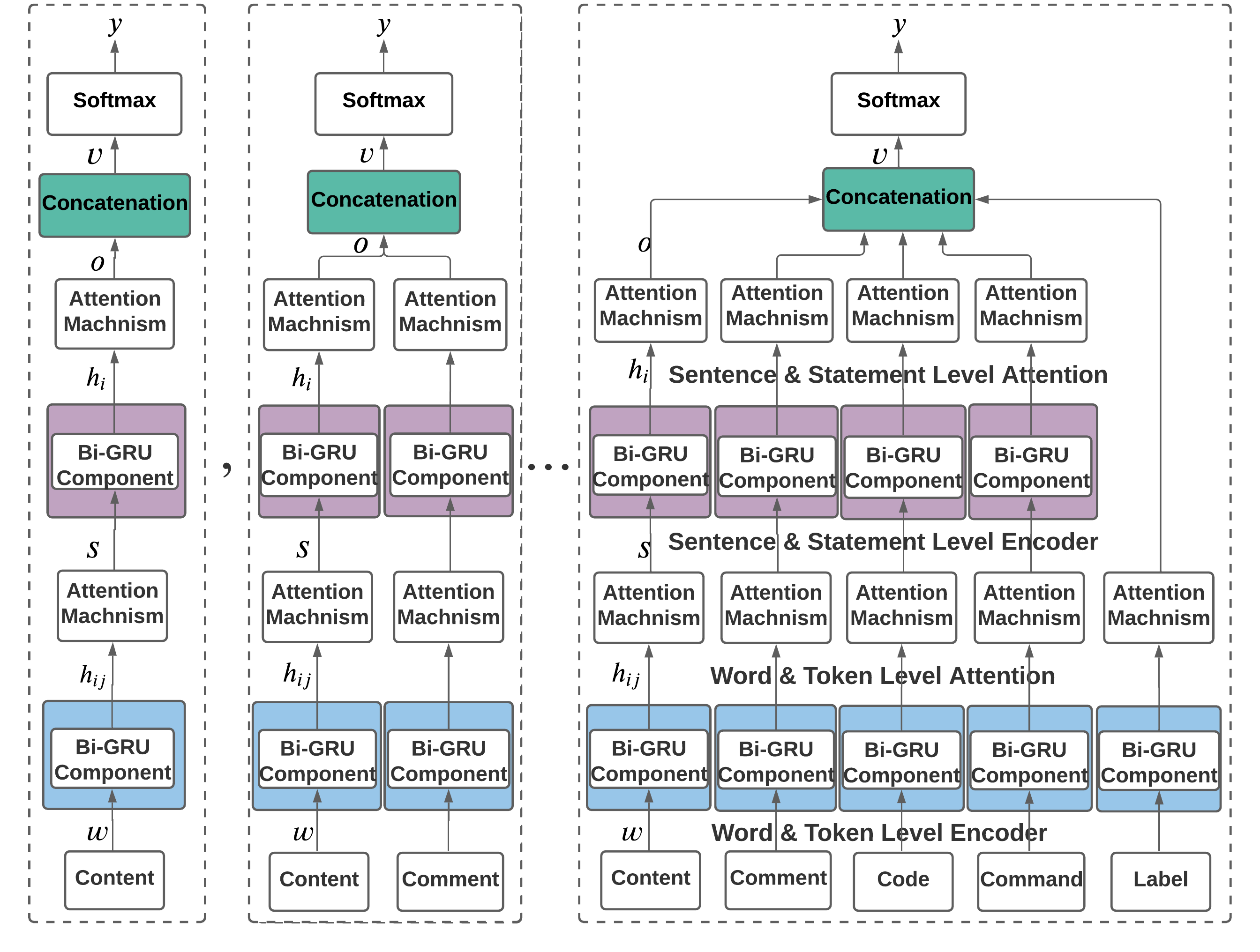}
  \caption{The MHAN structure with alternative combinations of feature types: from 1 up to 5 types of features.}
  \label{fig:structure}
\end{figure}

As shown in Figure~\ref{fig:structure}, we design a flexible structure of MHAN such that different feature types of the DL GitHub reports can be easily appended or removed. That is, MHAN can be built by simply considering one type of feature or anything up to all the five feature types we discussed in Section~\ref{sec:chall}. In Section~\ref{sec:rq2}, we will empirically identify the best combination of feature types for \approach.

Suppose that a feature has $L$ sentences/statements and each of them, denoted as $s_i$, contains $T_i$ words/tokens. $w_{ij}$ is the $j$th word/token in the $i$th sentence/statement and we have the corresponding word vector as $\mathbf{w_i}=\{w_{i1},w_{i2}, \cdots, w_{iT}\}$ for a sentence/statement of $T$ length. An example could be: \texttt{``the error is too high''} is treated as $\mathbf{w_i}=\{\textcolor{blue}{\texttt{`the'}}, \textcolor{blue}{\texttt{`error'}}, \textcolor{blue}{\texttt{`is'}}, \textcolor{blue}{\texttt{`too'}}, \textcolor{blue}{\texttt{`high'}}\}$. Note that for \textit{code} and \textit{command} features, we use tokenization to extract the tokens, e.g., code snippet \texttt{\textbf{\textcolor{blue}{model.compile(loss=\textcolor{red!70!black}{``crossentropy''})}}} can be represented as $\mathbf{w_i}=\{$\textcolor{blue}{\texttt{`model'}}, \textcolor{blue}{\texttt{`.'}}, \textcolor{blue}{\texttt{`compile'}}, \textcolor{blue}{\texttt{`('}}, \textcolor{blue}{\texttt{`loss'}}, \textcolor{blue}{\texttt{`='}}, \textcolor{blue}{\texttt{`$<$STR\_LIT$>$'}}, \textcolor{blue}{\texttt{`)'}}$\}$. In what follows, we will present the internal details of MHAN.

\subsection{Word and Token Level (First Hierarchy)} 

\subsubsection{Encoder} 

The first step in our MHAN is to embed the words/tokens of the sentences/statements in a feature type into numeric vectors for straightforward computation. In this work, we utilize \texttt{GloVe}~\cite{DBLP:conf/emnlp/PenningtonSM14} instead of the commonly used \texttt{word2vec}~\cite{DBLP:journals/corr/abs-1301-3781}, for two reasons: (1) it has fewer parameters; (2) it has been reported to be better on tasks with analogy and exploit the statistics of the corpus well~\cite{DBLP:conf/emnlp/PenningtonSM14}, which fits the nature of our problem.

Specifically, for a given feature type, \texttt{GloVe} embeds the vector of words/tokens for each sentence/statement, $\mathbf{w_i}$, into a sequence of numeric vector which we denote as $\mathbf{V_i}=\{\mathbf{v_{i1}},\mathbf{v_{i2}}, \cdots, \mathbf{v_{iT}}\}$, where $\mathbf{v_{ij}}$ is a vector that represents $w_{ij}$. Next, the embedded vector is fed into the bi-directional GRU (Bi-GRU) layer~\cite{DBLP:journals/corr/BahdanauCB14} to encode annotations of words/tokens by summarizing information from both directions for the sentence/statement, and hence
incorporate the contextual information. The purpose of this step is to build the state vector $\mathbf{{h_{ij}}}$ for each word/token $w_{ij}$, which can be expressed as:
\begin{equation}
\mathbf{h_{ij}}=\overrightarrow{\mathbf{h_{ij}}} \oplus \overleftarrow{\mathbf{h_{ij}}}=\overrightarrow{GRU}(\mathbf{v_{ij}}, \mathbf{V_i}) \oplus  \overleftarrow{GRU}(\mathbf{v_{ij}},\mathbf{V_i})
\label{eq:1}
\end{equation} 
whereby $\oplus$ is a concatenation operation. $\overrightarrow{\mathbf{h_{ij}}}$ denotes the annotation for $w_{ij}$ when reading from $w_{i1}$ to $w_{iT}$ by GRU, while $\overleftarrow{\mathbf{h_{ij}}}$ denotes that from $w_{iT}$ to $w_{i1}$. Therefore, $\mathbf{h_{ij}}$ is actually a concatenation of the state vectors that read from the opposed directions.

\subsubsection{Attention} 

Once we obtain the annotations of all words/tokens in a sentence/statement, MHAN exploits an attention mechanism to focus only on the most important ones, which is expressed as:
\begin{equation}
    \mathbf{u_{ij}}=\tanh\left(\mathbf{W_w}\mathbf{h_{ij}}+\mathbf{b_w}\right)
    \label{w:1}
\end{equation}
\begin{equation}
    \alpha_{ij\ }=\frac{\exp\left(\mathbf{u^\top_{ij}} \mathbf{u_w}\right)}{\sum_j\exp\left(\mathbf{u^\top_{ij}} \mathbf{u_w}\right)}
    \label{w:2}
\end{equation}
\begin{equation}
\mathbf{s_i}=\sum_j^{ }\alpha_{ij}\mathbf{h_{ij}}
\label{w:3}
\end{equation}
In Equation~(\ref{w:1}), the obtained states $\mathbf{{h_{ij}}}$ are used to train a one-layer Multilayer Perceptron (MLP), whereby $\mathbf{W_w}$ and $\mathbf{b_w}$ are weighted matrix and bias vector in that network, respectively. The output $\mathbf{{u_{ij}}}$ is taken as the hidden representation of $\mathbf{{h_{ij}}}$. The importance of each word/token, $\alpha_{ij}$, is measured as the outcome of a \textit{softmax} function~\cite{DBLP:conf/naacl/YangYDHSH16}, according to the normalized similarity between $\mathbf{{u_{ij}}}$ and $\mathbf{{u_{w}}}$, which is a word/token level context vector that can be jointly learned as part of the training~\cite{DBLP:conf/naacl/YangYDHSH16,sukhbaatar2015end,DBLP:conf/icml/KumarIOIBGZPS16}. Finally, the vector for sentence/statement, denoted as $\mathbf{s_i}$, is produced as a weighted-sum of all words/tokens for the next hierarchy level.

Recall the example from Table~\ref{tb:word-exp}, in this way, the words \texttt{\textcolor{blue}{`speed'}} and \texttt{\textcolor{blue}{`slowed'}} would get higher weights, as they are more semantically correlated to the contexts of \texttt{model} and \texttt{prediction}, which helps to identify DL \bugtype bug related reports. Similarly, from Figure~\ref{fig:code-exp}, the token \texttt{\textbf{\textcolor{blue}{time}}} would have higher weight in the contexts of \texttt{\textbf{\textcolor{blue}{Session}}} and \texttt{\textbf{\textcolor{blue}{run}}}, which is important to the identification.

\subsection{Sentence and Statement Level (Second Hierarchy)} 
\subsubsection{Encoder} 

For the \textit{label} feature, MHAN only allows for one hierarchy since it can be considered as a special case where a sentence has exactly one word. However, for the other feature types, more semantic information can be extracted at the sentence and statement level. Given the vector for sentence/statement $\mathbf{s_i}$, its annotation $\mathbf{h_{i}}$ can be encoded by using the bi-directional GRU:
\begin{equation}
\mathbf{h_{i}}=\overrightarrow{\mathbf{h_{i}}} \oplus \overleftarrow{\mathbf{h_{i}}}=\overrightarrow{GRU}(\mathbf{s_{i}}) \oplus  \overleftarrow{GRU}(\mathbf{s_{i}})
\end{equation} 
where all other notations are the same as Equation~\ref{eq:1}.

\subsubsection{Attention} 

Similar to the previous hierarchy, the sentences/statements that provided more information to the prediction are rewarded using the following:
\begin{equation}
    \mathbf{u_{i}}=\tanh\left(\mathbf{W_s}\mathbf{h_{i}}+\mathbf{b_s}\right)
    \label{s:1}
\end{equation}
\begin{equation}
    \alpha_{i\ }=\frac{\exp\left(\mathbf{u^\top_{i}} \mathbf{u_s}\right)}{\sum_i\exp\left(\mathbf{u^\top_{i}} \mathbf{u_s}\right)}
    \label{s:2}
\end{equation}
\begin{equation}
\mathbf{o}=\sum_i^{ }\alpha_{i}\mathbf{h_{i}}
\label{s:3}
\end{equation}
where $\mathbf{{u_{s}}}$ is a sentence/statement level context vector; $\mathbf{o}$ is the vector that represents the information of a given feature type, including the attentions to both word (token) and sentences (statements). In this way, the important sentence from Table~\ref{tb:word-exp} and the important statements from Figure~\ref{fig:code-exp} would contribute more in $\mathbf{o}$.

\subsection{Multifaceted Hierarchy} 

In MHAN, we design the above two levels of hierarchy for every feature type of GitHub reports we consider, except for the label, which only has the word level. As a result, the above hierarchical learning process is conducted for each feature type independently. Since according to Equation~\ref{s:3}, each feature would output a vector, representing the semantic information extracted (the \textit{label} feature would output a vector representation using Equation~\ref{w:3}), on top of that, we concatenate them to form the final vector $\mathbf{v}$: 
\begin{equation}
    \mathbf{v} = \oplus^k_{i=1} \mathbf{o_i}
\end{equation}
\begin{equation}
    y = \text{softmax}(\mathbf{v})
\end{equation}
where $k$ is the number of feature types ($1 \leq k \leq 5$) and $\oplus^k_{i=1}$ denotes repeated concatenation. For example, if the \textit{content}, \textit{comment}, and \textit{code} are considered in MHAN, then we have $\mathbf{v}=\mathbf{o_{content}} \oplus \mathbf{o_{comment}} \oplus \mathbf{o_{code}}$. Finally, $\mathbf{v}$ serves as the input to a \textit{softmax} function, which predicts $y$, indicating a given GitHub report for the DL frameworks is a report related to a \bugtype bug or not.
\section{\approach: an end-to-end tool}
\label{sec:end}

\begin{figure}[t!]
  \centering
  \includegraphics[width=\columnwidth]{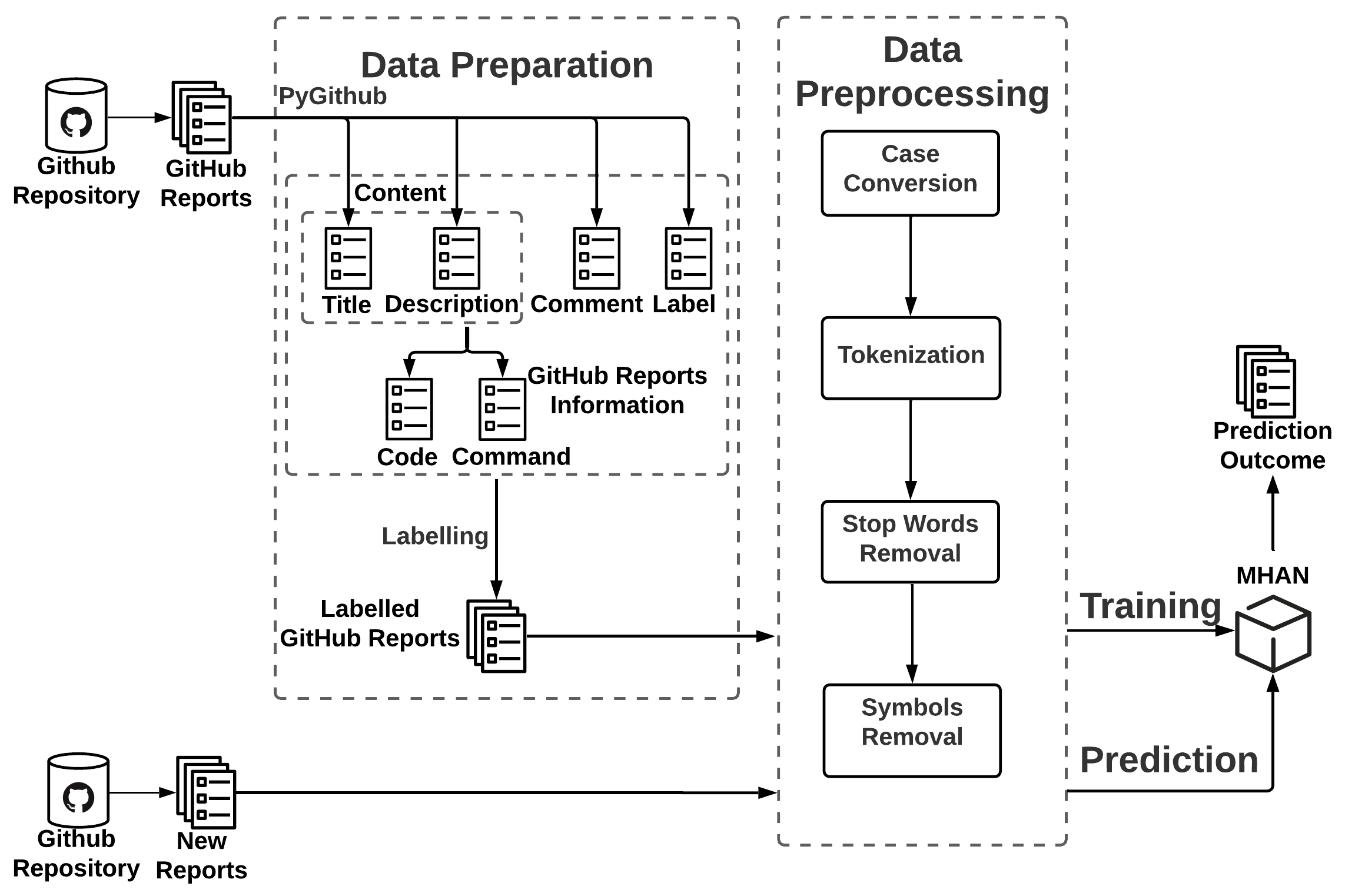}
  \caption{The architecture and deployment flow of \approach}
  \label{fig:end2end}
\end{figure}

\approach~was designed as an end-to-end tool that can be applied directly to a DL framework hosted on GitHub. As illustrated in Figure \ref{fig:end2end}, to deploy \approach, there are four key phases: \textit{data preparation}, \textit{data preprocessing}, \textit{training}, and \textit{prediction}, each of which is outlined below. Note that \approach only needs to train once (or as needed thereafter), and it can then identify newly submitted reports.

\subsection{Data Preparation}

In this work, we use the dataset as collected by Long and Chen~\cite{longchen2022} and we pick the top 5 DL frameworks with the largest number of reports. To further enrich the dataset, we extend the samples by including those reports that are not relevant to non-functional bugs following the same protocol. able~\ref{tab:dataset} summarizes the collected total of 3,712 GitHub reports for our experiments. They also allow us to conduct a pilot study that motivates some of the designs of \approach.

\begin{table}[t!]
\centering
  \caption{Characteristics of the collected GitHub reports. ($\mathcal{P}$ and $\mathcal{N}$ denote the number of \bugtype bug related reports and those are not \bugtype bug related, respectively; $\mathcal{P}$\% is the percentage of \bugtype bug related reports.)}
  \label{tab:dataset}
  \footnotesize

  \begin{tabular}{l c c c c}
    \toprule
    \textbf{Project}&\textbf{$\mathcal{P}$}&\textbf{$\mathcal{N}$} & \textbf{Total}&\textbf{$\mathcal{P}$\%}\\
    \midrule
    \rowcolor{white!20}\texttt{TensorFlow} & 279 & 1211 & 1490 & 18.7\%\\
    \texttt{PyTorch} & 95 & 657 & 752 & 12.6\%\\
    \rowcolor{white!20}\texttt{Keras} & 135 & 533 & 668 & 20.2\%\\
    \texttt{MXNet} & 65 & 451 & 516 & 12.6\%\\
    \rowcolor{white!20}\texttt{Caffe} & 33 & 253 & 286 & 11.5\%\\
    \midrule
    \textbf{TOTAL} & 607 & 3105 & 3712 & 16.4\%\\
  \bottomrule
\end{tabular}

\end{table}

\subsection{Data Preprocessing}

We preprocessed all collected reports via the steps below: 

\begin{enumerate}
    \item \textbf{Case Conversion:} All upper case texts are converted into lower case ones.
    \item  \textbf{Tokenization:} Sentences/statements are splitted into word tokens before embedding with \texttt{GloVe}. In this work, we use the Tokenizer\footnote{https://tinyurl.com/wbx7edeb} in \texttt{Keras}.
    \item  \textbf{Stop Words Removal:} Stop words, such as \texttt{\textcolor{blue}{``the''}} and \texttt{\textcolor{blue}{``is''}}, are removed. They appear frequently in natural language but with little contribution to semantic meaning.
    \item  \textbf{Symbols Removal:} HTML tags and punctuation marks that appear in the GitHub reports are also removed.
\end{enumerate}

\subsection{Training MHAN}

In \approach, the MHAN is trained using a stochastic algorithm called Adam~\cite{DBLP:journals/corr/KingmaB14} with $\beta_1=0.9$, $\beta_2=0.98$ and $\epsilon=10^{-9}$, as recommended by \citeauthor{DBLP:conf/nips/VaswaniSPUJGKP17}~\cite{DBLP:conf/nips/VaswaniSPUJGKP17}. For the structure of MHAN, we follow the values used by \citeauthor{DBLP:conf/naacl/YangYDHSH16}~\cite{DBLP:conf/naacl/YangYDHSH16}: a batch size of 64 with training epochs of 25; while the size of GRU layer and the embedding dimension are both 100. Since the report identification for \bugtype bugs in DL frameworks is a binary classification problem, we use the binary cross-entropy as the loss function in the training:
\begin{equation}
    \mathcal{L}=-\frac{1}{n}\sum_r^{}y_r\cdot\log y_r+\left(1-y_r\right)\cdot\log\left(1-y_r\right)
\end{equation}
where $n$ represents the number of training samples; $y_r$ denotes the prediction given the $i$th sample, where each sample refers to a GitHub report in this work.

\subsection{Prediction}

\approach~can benefit from those datasets as an initial starting point and be used in practice, since it learns the semantic information of GitHub reports; it can be further consolidated as more data is collected and labeled. Therefore, the training only happens once or as needed, but the learned knowledge can then be generalized for identifying new reports. This ensures the scalability of \approach.

\begin{table}[t!]
\centering
  \caption{Aliases for different combinations of feature types in \approach. \approach$_A$ essentially uses the classic HAN. The \textit{content} feature is handled by the best scheme from RQ1.}
  \label{tb:feature-type}
  \footnotesize
  \begin{tabular}{ll}
    \toprule
    \textbf{Alias}&\textbf{Feature Types}\\
    \midrule
    
    \approach$_A$&\textit{content}\\
    \approach$_B$&\textit{content}, \textit{comment}\\
    \approach$_C$&\textit{content}, \textit{code}\\
    \approach$_D$&\textit{content}, \textit{command}\\
    \approach$_E$&\textit{content}, \textit{label}\\
    \approach$_F$&\textit{content}, \textit{comment}, \textit{code}\\
    \approach$_G$&\textit{content}, \textit{comment}, \textit{command}\\
    \approach$_H$&\textit{content}, \textit{comment}, \textit{label}\\
    \approach$_I$&\textit{content}, \textit{code}, \textit{command}\\
    \approach$_J$&\textit{content}, \textit{code}, \textit{label}\\
    \approach$_K$&\textit{content}, \textit{command}, \textit{label}\\
    \approach$_L$&\textit{content}, \textit{comment}, \textit{command}, \textit{label}\\
    \approach$_M$&\textit{content}, \textit{comment}, \textit{command}, \textit{code}\\
    \approach$_N$&\textit{content}, \textit{comment}, \textit{code}, \textit{label}\\
    \approach$_O$&\textit{content}, \textit{command}, \textit{code}, \textit{label}\\
       \approach$_P$&\textit{content}, \textit{comment}, \textit{command}, \textit{code}, \textit{label}\\
  \bottomrule
 \end{tabular}

\end{table}

\section{Evaluation}
\label{sec:exp}
Here, we specify the research questions and settings.

\subsection{Research Questions}
\label{sec:rq}

We seek to answer four research questions (RQs):

\begin{itemize}
     \item \textbf{RQ1:} How \textit{title} and \textit{description} can be best handled as part of the \textit{content} feature?
     \item \textbf{RQ2:} What is the generally best combination of feature types for \approach?
     \item \textbf{RQ3:} How effective is \approach~against the state-of-the-arts in identifying \bugtype bug related reports for DL frameworks?
     \item \textbf{RQ4:} Why does \approach~work?
\end{itemize}

Since the \textit{title} and \textit{description} are both the fundamental sources of information in the \textit{content} feature type of a GitHub report (and are used in existing work~\cite{DBLP:journals/infsof/GomesTC21,DBLP:journals/ese/HerboldTT20,DBLP:journals/tr/UmerLI20,DBLP:conf/icmla/0003FWBL18,DBLP:conf/icsm/KallisSCP19}), we ask \textbf{RQ1} to study what is the best scheme to handle them. To that end, we compare three schemes:

\begin{itemize}
    \item \approach$_{title}$: only \textit{title} is considered.
    \item \approach$_{desc}$: only \textit{description} is considered.
    \item \approach$_{title+desc}$: both \textit{title} and \textit{description} are concatenated together as part of the \textit{content} feature. 

\end{itemize}

To eliminate unnecessary noise, we study \textbf{RQ1} by using the variant of \approach~where only the \textit{content} feature is considered.

We ask \textbf{RQ2} to understand whether all (or some) of the feature types identified in Section~\ref{sec:chall} can be useful for \approach. To this end, we compare the variants of \approach~under all possible combinations of the feature types combination, as shown in Table~\ref{tb:feature-type}. Since the \textit{content} is the most fundamental feature type, it is always considered, and we make use of the best variant from \textbf{RQ1} as the default treatment for the \textit{content} feature. In this way, \textbf{RQ2} also enables us to confirm whether the extended multifacetedness in the MHAN can be helpful, as we consider \approach~with only the \textit{content} feature (i.e., \approach$_A$), which essentially uses the
classic HAN. Note that, when a GitHub report does not contain data for certain types of features, e.g., one without any code, we build a dummy vector with only $0$ for such a feature type in the underlying MHAN of \approach.

    









\begin{table}
  \caption{The state-of-the-art classifiers compared. TF-IDF represents term frequency–inverse document frequency. Note that all classifiers were designed to consider title and description in their original work.}
  \label{tab:state}
  \footnotesize
      \setlength{\tabcolsep}{1mm}
\resizebox{\columnwidth}{!}{
  \begin{tabular}{llll}
    \toprule
    \textbf{Classifier}&\textbf{Description}&\textbf{Used By}&\textbf{Year}\\
    \midrule
    
    TF-IDF/$k$NN&TF-IDF with $k$-Nearest Neighbors.&\cite{DBLP:journals/infsof/GomesTC21,kukkar2018supervised}&2018-2021 \\

    TF-IDF/MLP&TF-IDF with MultiLayer Perceptrons.&\cite{DBLP:journals/infsof/GomesTC21}&2021 \\

    TF-IDF/NB&TF-IDF with Naive Bayes.&\cite{DBLP:journals/infsof/GomesTC21,jayagopal2021bug,gondaliya2018learning}&2018-2021  \\

    TF-IDF/RF&TF-IDF with Random Forest.&\cite{DBLP:journals/infsof/GomesTC21,jayagopal2021bug}&2021 \\

    TF-IDF/SVM&TF-IDF with Support Vector Machine.&\cite{DBLP:journals/infsof/GomesTC21,jayagopal2021bug,gondaliya2018learning}&2018-2021\\

    \texttt{TextCNN}& A convolutional neural network for text.&\cite{DBLP:journals/tr/UmerLI20}&2020\\

    LSTM & Long short-term memory networks.&\cite{DBLP:conf/icmla/0003FWBL18,DBLP:conf/internetware/QinS18,gondaliya2018learning}&2018 \\

    \texttt{fastText}& A text classification tool from Facebook.&\cite{DBLP:conf/icsm/KallisSCP19}&2019\\

    \texttt{auto-fastText} & \texttt{fastText} with auto-parameter tuning.&\cite{DBLP:journals/ese/HerboldTT20}&2020\\
  \bottomrule
\end{tabular}
}
\end{table}

To verify the effectiveness of \approach, in \textbf{RQ3}, we compare the best variant of \approach~(from \textbf{RQ2}) against nine state-of-the-art classifiers from the literature, including deep learning-based classifiers such as \texttt{LSTM} and \texttt{TextCNN}, as shown in Table~\ref{tab:state}.

\subsection{Experiment Procedure}

For \approach~(and variants) and other compared classifiers under a DL framework, our experiment follows the practice of bootstrapping (without replacement)~\cite{abney2002bootstrapping}, which consists of the steps below:

\begin{enumerate}
    \item Randomly split the dataset into 80\% for training and 20\% for the holdout testing.
    \item Train the classifiers with the training data. If a classifier under training comes from a state-of-the-art tool, conducting hyperparameter tuning.
    \item Evaluate the performance over the testing data.
    \item Repeat from step (1) for 30 runs.
\end{enumerate}

All the experiments in this work are implemented with Python 3.7 using the API from \texttt{Keras}, and run on a machine with 2.2GHz CPU, 12GB RAM, and a NVIDIA 16GB GPU .

\subsection{Metric}

In this work, the area under the receiver operating characteristic (ROC) curve, i.e., AUC~\cite{DBLP:journals/tkde/HuangL05}, is mainly applied as the performance metric to measure the effectiveness of a classifier over the holdout testing data. Because AUC does not require a particular threshold~\cite{ZhouYLCLZQX18}, which is difficult to tweak in order to carry out an unbiased assessment. Besides, it has been shown that the AUC is not sensitive to imbalanced data~\cite{LiJZ18,DBLP:journals/computing/KimCL17,DBLP:journals/prl/Fawcett06}, which is the case for this task (Table~\ref{tab:dataset}). Moreover, we also apply the precision, recall and F-measure to evaluate the performance of a classifier on difference aspects.

\begin{table*}[t]
    \caption{Scott-Knott rank and AUC on the schemes for title and description in \approach. ``Med'' denotes median AUC over 30 runs; \quartexp{0}{20}{10}{20} shows the 25\%, 50\%, and 75\% AUC percentiles. Rows are sorted from the best based on rank, median, and then IQR.}
    \label{tab:exp1}
  \begin{center}
    \begin{adjustbox}{max width = \textwidth}
     \footnotesize

    \begin{tabular}{c@{}c@{}c}
        \begin{tabular}{lcccl}
            \cellcolor[gray]{1}\textbf{Scheme} & \cellcolor[gray]{1}\textbf{Rank} &     \cellcolor[gray]{1}\textbf{Med} & \cellcolor[gray]{1}\textbf{IQR} & \cellcolor[gray]{1}\\
            \hline
            \approach$_{title+desc}$ & 1 & 0.854 & 0.036 & \quart{83.7}{3.6}{85.4}{100} \\
            \approach$_{title}$ & 2 & 0.841 & 0.051 & \quart{81.8}{5.1}{84.1}{100} \\
            \approach$_{desc}$ & 3 & 0.754 & 0.052 & \quart{72}{5.2}{75.4}{100}\\
        \end{tabular} & 
        \begin{tabular}{|lcccl}
           \cellcolor[gray]{1}\textbf{Scheme} & \cellcolor[gray]{1}\textbf{Rank} &     \cellcolor[gray]{1}\textbf{Med} & \cellcolor[gray]{1}\textbf{IQR} & \cellcolor[gray]{1}\\
            \hline
            \approach$_{title+desc}$ & 1 & 0.757 & 0.034 & \quart{73.6}{3.4}{75.7}{100} \\
            \approach$_{desc}$ & 2 & 0.697 & 0.032 & \quart{68.6}{3.2}{69.7}{100} \\
            \approach$_{title}$ & 3 & 0.683 & 0.051 &  \quart{65.7}{5.1}{68.3}{100} \\
        \end{tabular} &
        \begin{tabular}{|lcccl}
            \cellcolor[gray]{1}\textbf{Scheme} & \cellcolor[gray]{1}\textbf{Rank} &     \cellcolor[gray]{1}\textbf{Med} & \cellcolor[gray]{1}\textbf{IQR} & \cellcolor[gray]{1} \\
            \hline
            \approach$_{title+desc}$ & 1 & 0.815 & 0.067 &  \quart{65.7}{15.1}{81.5}{100}\\
            \approach$_{title}$ & 2 & 0.793 & 0.327 & \quart{50}{32.7}{79.3}{100} \\
            \approach$_{desc}$ & 3 & 0.701 & 0.131 &  \quart{62.6}{13.1}{70.1}{100} \\
        \end{tabular} \\
        \textbf{(a). \texttt{TensorFlow}} & \textbf{(b). \texttt{PyTorch}} & \textbf{(c). \texttt{Keras}}
        \\
        \\
        \begin{tabular}{lcccl}
             \cellcolor[gray]{1}\textbf{Scheme} & \cellcolor[gray]{1}\textbf{Rank} &     \cellcolor[gray]{1}\textbf{Med} & \cellcolor[gray]{1}\textbf{IQR} & \cellcolor[gray]{1}
            \\
            \hline
            \approach$_{title+desc}$ & 1 & 0.744 & 0.350 & \quart{50}{35}{74.4}{100} \\
            \approach$_{desc}$ & 2 & 0.648 & 0.220 & \quart{50}{22}{64.8}{100} \\
            \approach$_{title}$ & 3 & 0.500 & 0.294 & \quart{50}{29.4}{50}{100} \\
        \end{tabular} &

        \begin{tabular}{|lcccl|}
            \cellcolor[gray]{1}\textbf{Scheme} & \cellcolor[gray]{1}\textbf{Rank} &     \cellcolor[gray]{1}\textbf{Med} & \cellcolor[gray]{1}\textbf{IQR} & \cellcolor[gray]{1}\\
            \hline
            \approach$_{title+desc}$ & 1 & 0.652 & 0.228 &  \quart{50}{22.8}{65.2}{100}\\
            \approach$_{desc}$ & 2 & 0.536 & 0.137 & \quart{50}{13.7}{53.6}{100} \\
            \approach$_{title}$ & 3 & 0.500 & 0.094 & \quart{50}{9.4}{50}{100}\\
        \end{tabular} &
        \begin{tabular}{lcccl}
            \cellcolor[gray]{1}\textbf{Scheme} & \cellcolor[gray]{1}\textbf{Rank} &     \hphantom{\cellcolor[gray]{1}\textbf{Med}} & \hphantom{\cellcolor[gray]{1}\textbf{IQR}} & \hphantom{\cellcolor[gray]{1}}\\
            \hline
            \approach$_{title+desc}$&5&\hphantom{0.000}&\hphantom{0.000}&\hphantom{\quart{50}{9.4}{50}{100}} \\
            \approach$_{desc}$ & 12&\hphantom{0.000}&\hphantom{0.000}&\hphantom{\quart{50}{9.4}{50}{100}} \\
            \approach$_{title}$ & 13&\hphantom{0.000}&\hphantom{0.000}&\hphantom{\quart{50}{9.4}{50}{100}} \\
        \end{tabular}\\
        \textbf{(d). \texttt{MXNet}}  & \textbf{(e). \texttt{Caffe}} & \textbf{(f). Total rank}\\
    \end{tabular}
  \end{adjustbox}
   \end{center}
\end{table*}

\subsection{Statistical Test}

To ensure the comparisons are statistically meaningful, we apply Scott-Knott test~\cite{DBLP:journals/tse/MittasA13} on all comparisons of the AUC over 30 runs. As a widely-used test in Software Engineering~\cite{10.1145/3514233}, Scott-Knott sorts the list of treatments (the classifiers) by their median AUC values. Next, it splits the list into two sub-lists with the largest expected difference~\cite{xia2018hyperparameter}. Formally, Scott-Knott test aims to find the best split by maximizing the difference $\Delta$ in the expected mean before and after each split:
\begin{equation}
    \Delta = \frac{|l_1|}{|l|}(\overline{l_1} - \overline{l})^2 + \frac{|l_2|}{|l|}(\overline{l_2} - \overline{l})^2
\end{equation}
whereby $|l_1|$ and $|l_2|$ are the sizes of two sub-lists ($l_1$ and $l_2$) from list $l$ with a size $|l|$. $\overline{l_1}$, $\overline{l_2}$, and $\overline{l}$ denote their mean AUC.

During the splitting, we apply a statistical hypothesis test $H$ to check if $l_1$ and $l_2$ are significantly different. This is done by using bootstrapping and $\hat{A}_{12}$~\cite{vargha2000critique} (a non-parametric effect size metric). If that is the case, Scott-Knott recurses on the splits. In other words, we divide the classifiers into different sub-lists if both bootstrap sampling and effect size test suggest that a split is statistically significant (with a confidence level of 99\%) and not a small effect $\hat{A}_{12} \geq 0.6$. The sub-lists are then ranked based on their mean AUC.

Therefore, when we say $A$ is better than $B$, their difference is indeed statistically significant.

\begin{table*}[ht!]
    \caption{Scott-Knott rank and AUC on the variants of \approach~(with the best scheme from RQ1). \setlength{\fboxsep}{1.5pt}\colorbox{blue!10}{blue} highlights the overall best variant. Formats are the same as Table~\ref{tab:exp1}.}
    \label{tab:exp2}
  \begin{center}
    \begin{adjustbox}{max width = \textwidth}
     
    \footnotesize
    \begin{tabular}{c@{}c@{}c}
        \begin{tabular}{lcccl}
            \cellcolor[gray]{1}\textbf{Variant} & \cellcolor[gray]{1}\textbf{Rank} &     \cellcolor[gray]{1}\textbf{Med} & \cellcolor[gray]{1}\textbf{IQR} & \cellcolor[gray]{1}\\
            \hline
            \approach$_H$ & 1 & 0.859 & 0.036 & \quart{83.8}{3.6}{85.9}{100} \\
            \cellcolor{blue!10}\approach$_F$ &\cellcolor{blue!10}1 &\cellcolor{blue!10}0.857 &\cellcolor{blue!10}0.025 &\cellcolor{blue!10}\quart{84.5}{2.5}{85.7}{100} \\
            \approach$_B$ & 1 & 0.853 & 0.040 & \quart{83.4}{4}{85.3}{100} \\
            \approach$_E$ & 2 & 0.858 & 0.045 & \quart{82.9}{4.5}{85.8}{100} \\
            \approach$_P$ & 2 & 0.857 & 0.042 & \quart{82.9}{4.2}{85.7}{100} \\
            \approach$_A$ & 2 & 0.854 & 0.036 & \quart{83.7}{3.6}{85.4}{100} \\
            \approach$_J$ & 2 & 0.854 & 0.043 & \quart{82.6}{4.3}{85.4}{100} \\
            \approach$_M$ & 2 & 0.849 & 0.035 & \quart{83.9}{3.5}{84.9}{100} \\
            \approach$_C$ & 2 & 0.845 & 0.040 & \quart{82.8}{4}{84.5}{100} \\
            \approach$_G$ & 3 & 0.852 & 0.035 & \quart{82.9}{3.5}{85.2}{100} \\
            \approach$_L$ & 3 & 0.847 & 0.044 & \quart{82.2}{4.4}{84.7}{100}\\
            \approach$_D$ & 3 & 0.846 & 0.031 & \quart{82.6}{3.1}{84.6}{100} \\
            \approach$_O$ & 3 & 0.845 & 0.025 & \quart{83.1}{2.5}{84.5}{100} \\
            \approach$_K$ & 3 & 0.845 & 0.061 & \quart{82}{6.1}{84.5}{100}  \\
            \approach$_N$ & 3 & 0.843 & 0.050 & \quart{81.9}{5}{84.3}{100} \\
            \approach$_I$ & 3 & 0.842 & 0.041 & \quart{83.1}{4.1}{84.2}{100} \\

        \end{tabular} & 
        \begin{tabular}{|lcccl}
           \cellcolor[gray]{1}\textbf{Variant} & \cellcolor[gray]{1}\textbf{Rank} &     \cellcolor[gray]{1}\textbf{Med} & \cellcolor[gray]{1}\textbf{IQR} & \cellcolor[gray]{1}\\
            \hline
            \approach$_A$ & 1 & 0.757 & 0.034 &  \quart{73.6}{3.4}{75.7}{100}\\
            \cellcolor{blue!10}\approach$_F$ & \cellcolor{blue!10}2 & \cellcolor{blue!10}0.748 & \cellcolor{blue!10}0.046 & \cellcolor{blue!10}\quart{72.6}{4.6}{74.8}{100} \\
            \approach$_J$ & 3 & 0.744 & 0.029 &  \quart{72.7}{2.9}{74.4}{100}\\
            \approach$_O$ & 3 & 0.742 & 0.059 & \quart{71.2}{5.9}{74.2}{100} \\
            \approach$_K$ & 3 & 0.740 & 0.030 & \quart{72.3}{3}{74}{100} \\
            \approach$_D$ & 3 & 0.735 & 0.039 & \quart{72.1}{3.9}{73.5}{100} \\
            \approach$_E$ & 4 & 0.733 & 0.047 & \quart{71.3}{4.7}{73.3}{100} \\
            \approach$_N$ & 4 & 0.726 & 0.034 & \quart{71.6}{3.4}{72.6}{100} \\
            \approach$_L$ & 4 & 0.726 & 0.034 & \quart{71.6}{3.4}{72.6}{100} \\
            \approach$_C$ & 5 & 0.723 & 0.054 & \quart{69.3}{5.4}{72.3}{100} \\
            \approach$_I$ & 5 & 0.718 & 0.043 & \quart{69.7}{4.3}{71.8}{100} \\
            \approach$_H$ & 6 & 0.720 & 0.035 & \quart{69.6}{3.5}{72}{100} \\
            \approach$_P$ & 6 & 0.714 & 0.040 & \quart{68.7}{4}{71.4}{100} \\
            \approach$_M$ & 6 & 0.711 & 0.032 & \quart{69.2}{3.2}{71.1}{100} \\
            \approach$_G$ & 7 & 0.696 & 0.034 & \quart{68.8}{3.4}{69.6}{100} \\
            \approach$_B$ & 8 & 0.681 & 0.033 & \quart{66.4}{3.3}{68.1}{100} \\
        \end{tabular} &
        \begin{tabular}{|lcccl}
            \cellcolor[gray]{1}\textbf{Variant} & \cellcolor[gray]{1}\textbf{Rank} &     \cellcolor[gray]{1}\textbf{Med} & \cellcolor[gray]{1}\textbf{IQR} & \cellcolor[gray]{1} \\
            \hline
            \approach$_C$ & 1 & 0.823 & 0.064 & \quart{80}{6.4}{82.3}{100} \\
            \approach$_O$ & 2 & 0.843 & 0.079 & \quart{79.1}{7.9}{84.3}{100} \\
            \approach$_J$ & 2 & 0.838 & 0.044 &  \quart{80.8}{4.4}{83.8}{100}\\
            \approach$_N$ & 2 & 0.835 & 0.076 & \quart{78.8}{7.6}{83.5}{100} \\
            \cellcolor{blue!10}\approach$_F$ & \cellcolor{blue!10}2 & \cellcolor{blue!10}0.828 & \cellcolor{blue!10}0.093 & \cellcolor{blue!10}\quart{77.4}{9.3}{82.8}{100} \\
            \approach$_P$ & 2 & 0.827 & 0.103 & \quart{77.2}{10.3}{82.7}{100} \\
            \approach$_B$ & 2 & 0.824 & 0.061 & \quart{79.8}{6.1}{82.4}{100} \\
            \approach$_D$ & 2 & 0.824 & 0.078 & \quart{77.3}{7.8}{82.4}{100} \\
            \approach$_L$ & 2 & 0.813 & 0.066 & \quart{78.3}{6.6}{81.3}{100} \\
            \approach$_G$ & 3 & 0.825 & 0.075 & \quart{78.3}{7.5}{82.5}{100} \\
            \approach$_I$ & 3 & 0.824 & 0.087 & \quart{75.8}{8.7}{82.4}{100} \\
            \approach$_E$ & 3 & 0.821 & 0.099 & \quart{76}{9.9}{82.1}{100} \\
            \approach$_M$ & 3 & 0.818 & 0.077 & \quart{77.2}{7.7}{81.8}{100} \\
            \approach$_A$ & 3 & 0.815 & 0.067 & \quart{77.8}{6.7}{81.5}{100} \\
            \approach$_K$ & 4 & 0.818 & 0.127 & \quart{73.9}{12.7}{81.8}{100} \\
            \approach$_H$ & 4 & 0.815 & 0.067 & \quart{77.6}{6.7}{81.5}{100} \\
        \end{tabular} \\
        \textbf{(a). \texttt{TensorFlow}} & \textbf{(b). \texttt{PyTorch}} & \textbf{(c). \texttt{Keras}}
        \\
        \\
        \begin{tabular}{lcccl}
             \cellcolor[gray]{1}\textbf{Variant} & \cellcolor[gray]{1}\textbf{Rank} &     \cellcolor[gray]{1}\textbf{Med} & \cellcolor[gray]{1}\textbf{IQR} & \cellcolor[gray]{1}
            \\
            \hline
            \approach$_M$ & 1 & 0.815 & 0.143 & \quart{72}{14.3}{81.5}{100} \\
            \approach$_J$ & 1 & 0.811 & 0.349 &  \quart{50.9}{34.9}{81.1}{100}\\
            \approach$_G$ & 1 & 0.793 & 0.183 &  \quart{69.1}{18.3}{79.3}{100}\\
            \approach$_K$ & 1 & 0.788 & 0.194 & \quart{67.7}{19.4}{78.8}{100} \\
            \approach$_O$ & 1 & 0.787 & 0.158 & \quart{69.6}{15.8}{78.7}{100} \\
            \approach$_L$ & 1 & 0.782 & 0.348 & \quart{55}{34.8}{78.2}{100} \\
            \cellcolor{blue!10}\approach$_F$ & \cellcolor{blue!10}1 & \cellcolor{blue!10}0.778 & \cellcolor{blue!10}0.199 & \cellcolor{blue!10}\quart{67.3}{19.9}{77.8}{100} \\
            \approach$_H$ & 1 & 0.768 & 0.187 & \quart{67.6}{18.7}{76.8}{100} \\
            \approach$_I$ & 2 & 0.793 & 0.355 & \quart{50}{35.5}{79.3}{100} \\
            \approach$_E$ & 2 & 0.788 & 0.349 & \quart{50}{34.9}{78.8}{100} \\
            \approach$_C$ & 2 & 0.775 & 0.251 & \quart{60.4}{25.1}{77.5}{100} \\
            \approach$_D$ & 2 & 0.769 & 0.308 & \quart{56.2}{30.8}{76.9}{100} \\
            \approach$_N$ & 2 & 0.761 & 0.371 & \quart{50}{37.1}{76.1}{100} \\
            \approach$_P$ & 2 & 0.750 & 0.232 & \quart{61.3}{23.2}{75}{100} \\
            \approach$_B$ & 3 & 0.768 & 0.339 & \quart{50}{33.9}{76.8}{100} \\
            \approach$_A$ & 3 & 0.74 & 0.350 & \quart{50}{35}{74.4}{100} \\
        \end{tabular} &

        \begin{tabular}{|lcccl|}
            \cellcolor[gray]{1}\textbf{Variant} & \cellcolor[gray]{1}\textbf{Rank} &     \cellcolor[gray]{1}\textbf{Med} & \cellcolor[gray]{1}\textbf{IQR} & \cellcolor[gray]{1}\\
            \hline
            \approach$_A$ & 1 & 0.652 & 0.228 & \quart{50}{22.8}{65.2}{100} \\
            \approach$_B$ & 1 & 0.652 & 0.247 & \quart{50}{24.7}{65.2}{100} \\
            \cellcolor{blue!10}\approach$_F$ & \cellcolor{blue!10}1 & \cellcolor{blue!10}0.646 & \cellcolor{blue!10}0.247 & \cellcolor{blue!10}\quart{50}{24.7}{64.6}{100}\\
            \approach$_C$ & 1 & 0.640 & 0.212 & \quart{50}{21.2}{64}{100} \\
            \approach$_E$ & 1 & 0.640 & 0.259 & \quart{50}{25.9}{64}{100} \\
            \approach$_D$ & 1 & 0.635 & 0.210 & \quart{50}{21}{63.5}{100} \\
            \approach$_I$ & 1 & 0.629 & 0.276 & \quart{50}{27.6}{62.9}{100} \\
            \approach$_K$ & 1 & 0.626 & 0.240 & \quart{50}{24}{62.6}{100} \\
            \approach$_H$ & 1 & 0.623 & 0.261 & \quart{50}{26.1}{62.3}{100} \\
            \approach$_O$ & 1 & 0.600 & 0.243 & \quart{50}{24.3}{60}{100} \\
            \approach$_G$ & 2 & 0.605 & 0.194 & \quart{50}{19.4}{60.5}{100} \\
            \approach$_L$ & 2 & 0.599 & 0.213 &  \quart{50}{21.3}{59.9}{100}\\
            \approach$_J$ & 2 & 0.531 & 0.228 &  \quart{50}{22.8}{53.1}{100}\\
            \approach$_P$ & 3 & 0.616 & 0.195 &  \quart{50}{19.5}{61.6}{100}\\
            \approach$_M$ & 3 & 0.553 & 0.168 & \quart{50}{16.8}{55.3}{100} \\
            \approach$_N$ & 3 & 0.553 & 0.245 & \quart{50}{24.5}{55.3}{100} \\
        \end{tabular} &
        \begin{tabular}{lcccl}
             \cellcolor[gray]{1}\textbf{Variant} & \cellcolor[gray]{1}\textbf{Rank} &     \hphantom{\cellcolor[gray]{1}\textbf{Med}} & \hphantom{\cellcolor[gray]{1}\textbf{IQR}} & \hphantom{\cellcolor[gray]{1}}\\
            \hline
            \cellcolor{blue!10}\approach$_F$ &\cellcolor{blue!10}7&\hphantom{0.000}&\hphantom{0.000}&\hphantom{\quart{50}{9.4}{50}{100}}  \\
            \approach$_O$ & 10&\hphantom{0.000}&\hphantom{0.000}&\hphantom{\quart{50}{9.4}{50}{100}}  \\
            \approach$_J$ & 10&\hphantom{0.000}&\hphantom{0.000}&\hphantom{\quart{50}{9.4}{50}{100}}\\
            \approach$_A$ & 10&\hphantom{0.000}&\hphantom{0.000}&\hphantom{\quart{50}{9.4}{50}{100}}\\
            \approach$_C$ & 11&\hphantom{0.000}&\hphantom{0.000}&\hphantom{\quart{50}{9.4}{50}{100}}  \\
            \approach$_D$ & 11&\hphantom{0.000}&\hphantom{0.000}&\hphantom{\quart{50}{9.4}{50}{100}}  \\
            \approach$_K$ & 12&\hphantom{0.000}&\hphantom{0.000}&\hphantom{\quart{50}{9.4}{50}{100}}  \\
            \approach$_L$ & 12&\hphantom{0.000}&\hphantom{0.000}&\hphantom{\quart{50}{9.4}{50}{100}}  \\
            \approach$_E$ & 12&\hphantom{0.000}&\hphantom{0.000}&\hphantom{\quart{50}{9.4}{50}{100}}  \\
            \approach$_H$ & 13&\hphantom{0.000}&\hphantom{0.000}&\hphantom{\quart{50}{9.4}{50}{100}}  \\
            \approach$_I$ & 14&\hphantom{0.000}&\hphantom{0.000}&\hphantom{\quart{50}{9.4}{50}{100}}  \\
            \approach$_N$ & 14&\hphantom{0.000}&\hphantom{0.000}&\hphantom{\quart{50}{9.4}{50}{100}}  \\
            \approach$_B$ & 15&\hphantom{0.000}&\hphantom{0.000}&\hphantom{\quart{50}{9.4}{50}{100}}  \\
            \approach$_M$ & 15&\hphantom{0.000}&\hphantom{0.000}&\hphantom{\quart{50}{9.4}{50}{100}}  \\
            \approach$_P$ & 15&\hphantom{0.000}&\hphantom{0.000}&\hphantom{\quart{50}{9.4}{50}{100}}  \\
            \approach$_G$ & 16&\hphantom{0.000}&\hphantom{0.000}&\hphantom{\quart{50}{9.4}{50}{100}}  \\
        \end{tabular}\\
        \textbf{(d). \texttt{MXNet}}  & \textbf{(e). \texttt{Caffe}} & \textbf{(f). Total rank}\\
    \end{tabular}
  \end{adjustbox}
   \end{center}
   \vspace{-0.1cm}
\end{table*}

\section{Results}
\label{sec:result}
We now present and discuss the experimental results.


\subsection{Schemes for Title and Description}

To answer \textbf{RQ1}, Table~\ref{tab:exp1} shows the results and statistics when comparing the three schemes mentioned in Section~\ref{sec:rq}. Clearly, we see that \approach$_{title+desc}$ significantly outperforms the other two by being consistently ranked as the sole best over the DL frameworks. In particular, the magnitudes of improvements are considerably high, as we can see from the AUC distributions.

An interesting finding is that, when considering only \textit{description} (\approach$_{desc}$), the performance drops significantly. This means that even the \textit{title} is generally a much shorter piece of text than the \textit{description}, it is important to take it into account when identifying \bugtype bug related reports for DL frameworks. This makes sense, as the title can often provide some of the most important words that can help \approach~to learn the overall semantic information.

Therefore, for \textbf{RQ1}, we say:

\begin{quotebox}
   \noindent
   \textit{\textbf{\underline{To RQ1:}} Concatenating title and description in the content is the most effective scheme for \approach.}
\end{quotebox}


\subsection{Variants of \approach}
\label{sec:rq2}

Table~\ref{tab:exp2} illustrates the results for \textbf{RQ2}, from which we can see that some of the variants perform much better than the others. Overall, the variant \approach$_F$, which considers the feature \textit{content}, \textit{comment}, and \textit{code} is the most promising one, suggesting that those feature types, when being independently considered together, are more likely to offer unique information for learning semantic knowledge. In particular, \approach$_F$ has been ranked the 1st for three DL frameworks and 2nd for two, leading to a total rank of 7. It is interesting to see that, as opposed to our initial intuition, using all five types of feature perform badly, producing a total rank of 15. This is indeed possible, because certain feature types may often contribute to a considerable amount of redundant information, leaving the useful knowledge blurred and hence increasing the difficulty of learning. Hence, we say:

\begin{quotebox}
   \noindent
   \textit{\textbf{\underline{To RQ2:}} Using \textit{content}, \textit{comment}, and \textit{code} (\approach$_F$) are generally the most promising feature combination for \approach. The extended multifacetedness in MHAN of \approach~can indeed help to considerably improve the result.}
\end{quotebox}


\begin{table*}[t!]
      \caption{Scott-Knott rank and AUC on comparing \approach~(the best variant from \textbf{RQ2}) with the state-of-the-art classifiers. The rows for \approach~are highlighted in \setlength{\fboxsep}{1.5pt}\colorbox{blue!10}{blue}. Formats are the same as Table~\ref{tab:exp1}.}
    \label{tab:exp3}
  \begin{center}
    \begin{adjustbox}{max width = \textwidth}
     
\footnotesize
    \begin{tabular}{c@{}c@{}c}
        \begin{tabular}{lcccl}
            \cellcolor[gray]{1}\textbf{Classifier} & \cellcolor[gray]{1}\textbf{Rank} &     \cellcolor[gray]{1}\textbf{Med} & \cellcolor[gray]{1}\textbf{IQR} & \cellcolor[gray]{1}\\
            \hline
            \cellcolor{blue!10}\approach & \cellcolor{blue!10}1 &\cellcolor{blue!10} 0.857 & \cellcolor{blue!10}0.025 & \cellcolor{blue!10} \quart{84.5}{2.5}{85.7}{100}\\
            TF-IDF/MLP & 2 & 0.790 & 0.023 & \quart{78.1}{2.3}{79}{100} \\
            \texttt{auto-fastText} & 3 & 0.786 & 0.039 & \quart{76.1}{3.9}{78.6}{100} \\
            TF-IDF/NB & 4 & 0.771 & 0.035 & \quart{74.8}{3.5}{77.1}{100} \\
            \texttt{TextCNN} & 5 & 0.700 & 0.070 & \quart{65.6}{7}{70}{100} \\
            LSTM & 6 & 0.681 & 0.102 & \quart{60.7}{10.2}{68.1}{100} \\
            TF-IDF/RF & 7 & 0.653 & 0.043 & \quart{62.7}{4.3}{65.3}{100} \\
            TF-IDF/KNN & 7 & 0.619 & 0.053 & \quart{59.6}{5.3}{61.9}{100} \\
            TF-IDF/SVM & 8 & 0.500 & 0.148 & \quart{50}{14.8}{50}{100} \\
            \texttt{fastText}& 9 & 0.500 & 0.000 & \quart{50}{0}{50}{100} \\
        \end{tabular} & 
        \begin{tabular}{|lcccl}
           \cellcolor[gray]{1}\textbf{Classifier} & \cellcolor[gray]{1}\textbf{Rank} &     \cellcolor[gray]{1}\textbf{Med} & \cellcolor[gray]{1}\textbf{IQR} & \cellcolor[gray]{1}\\
            \hline
            \cellcolor{blue!10} \approach & \cellcolor{blue!10}1 & \cellcolor{blue!10}0.748 & \cellcolor{blue!10}0.046 & \cellcolor{blue!10}\quart{72.6}{4.6}{74.8}{100}  \\
            TF-IDF/NB & 2 & 0.725 & 0.041 & \quart{70.7}{4.1}{72.5}{100} \\
            TF-IDF/MLP & 3 & 0.640 & 0.108 & \quart{58.8}{10.8}{64}{100} \\
            \texttt{auto-fastText} & 4 & 0.557 & 0.108 & \quart{50.4}{10.8}{55.7}{100} \\
            TF-IDF/SVM & 4 & 0.500 & 0.150 & \quart{50}{15}{50}{100} \\
            TF-IDF/RF & 5 & 0.539 & 0.035 & \quart{52}{3.5}{53.9}{100} \\
            \texttt{TextCNN} & 5 & 0.518 & 0.077 & \quart{50}{7.7}{51.8}{100} \\
            TF-IDF/KNN & 6 & 0.500 & 0.027 & \quart{50}{2.7}{50}{100} \\
            LSTM & 7 & 0.500 & 0.000 & \quart{50}{0}{50}{100} \\
            \texttt{fastText}& 8 & 0.500 & 0.000 & \quart{50}{0}{50}{100} \\
        \end{tabular} &
        \begin{tabular}{|lcccl}
            \cellcolor[gray]{1}\textbf{Classifier} & \cellcolor[gray]{1}\textbf{Rank} &     \cellcolor[gray]{1}\textbf{Med} & \cellcolor[gray]{1}\textbf{IQR} & \cellcolor[gray]{1} \\
            \hline
             \cellcolor{blue!10}\approach & \cellcolor{blue!10}1 & \cellcolor{blue!10}0.828 & \cellcolor{blue!10}0.093 & \cellcolor{blue!10}\quart{77.4}{9.3}{82.8}{100} \\
            TF-IDF/NB & 2 & 0.781 & 0.054 & \quart{74.9}{5.4}{78.1}{100} \\
            TF-IDF/MLP & 3 & 0.714 & 0.069 & \quart{69}{6.9}{71.4}{100} \\
            \texttt{auto-fastText} & 4 & 0.708 & 0.058 & \quart{69}{5.8}{70.8}{100} \\
            TF-IDF/RF & 5 & 0.643 & 0.066 & \quart{60.4}{6.6}{64.3}{100} \\
            TF-IDF/SVM & 6 & 0.611 & 0.090 & \quart{57.7}{9}{61.1}{100} \\
            \texttt{TextCNN} & 7 & 0.617 & 0.057 & \quart{58.2}{5.7}{61.7}{100} \\
            LSTM & 7 & 0.592 & 0.121 & \quart{53.5}{12.1}{59.2}{100} \\
            TF-IDF/KNN & 8 & 0.582 & 0.058 & \quart{56.7}{5.8}{58.2}{100} \\
            \texttt{fastText}& 9 & 0.500 & 0.000 & \quart{50}{0}{50}{100} \\
        \end{tabular} \\
         \textbf{(a). \texttt{TensorFlow}} & \textbf{(b). \texttt{PyTorch}} & \textbf{(c). \texttt{Keras}}
        \\
        \\
        \begin{tabular}{lcccl}
             \cellcolor[gray]{1}\textbf{Classifier} & \cellcolor[gray]{1}\textbf{Rank} &     \cellcolor[gray]{1}\textbf{Med} & \cellcolor[gray]{1}\textbf{IQR} & \cellcolor[gray]{1}
            \\
            \hline
            \cellcolor{blue!10}\approach & \cellcolor{blue!10}1 & \cellcolor{blue!10}0.778 & \cellcolor{blue!10}0.199 & \cellcolor{blue!10}\quart{67.3}{19.9}{77.8}{100} \\
            TF-IDF/NB & 1 & 0.770 & 0.051 & \quart{73.6}{5.1}{77}{100} \\
            TF-IDF/MLP & 2 & 0.659 & 0.078 & \quart{61}{7.8}{65.9}{100} \\
            \texttt{auto-fastText} & 3 & 0.616 & 0.085 & \quart{58.8}{8.5}{61.6}{100} \\
            TF-IDF/KNN & 4 & 0.538 & 0.070 & \quart{50.6}{7}{53.8}{100} \\
            TF-IDF/SVM & 4 & 0.500 & 0.099 & \quart{50}{9.9}{50}{100} \\
            TF-IDF/RF & 5 & 0.500 & 0.034 & \quart{50}{3.4}{50}{100} \\
            \texttt{TextCNN} & 5 & 0.500 & 0.000 & \quart{50}{0}{50}{100} \\
            LSTM & 5 & 0.500 & 0.000 & \quart{50}{0}{50}{100} \\
            \texttt{fastText}& 6 & 0.500 & 0.000 & \quart{50}{0}{50}{100} \\
        \end{tabular} &

        \begin{tabular}{|lcccl|}
            \cellcolor[gray]{1}\textbf{Classifier} & \cellcolor[gray]{1}\textbf{Rank} &     \cellcolor[gray]{1}\textbf{Med} & \cellcolor[gray]{1}\textbf{IQR} & \cellcolor[gray]{1}\\
            \hline
            TF-IDF/NB & 1 & 0.665 & 0.060 & \quart{64.7}{6}{66.5}{100} \\
            \cellcolor{blue!10}\approach & \cellcolor{blue!10}2 & \cellcolor{blue!10}0.646 & \cellcolor{blue!10}0.247 & \cellcolor{blue!10}\quart{50}{24.7}{64.6}{100} \\
            TF-IDF/MLP & 3 & 0.590 & 0.102 & \quart{56.5}{10.2}{59}{100} \\
            \texttt{auto-fastText} & 4 & 0.586 & 0.068 & \quart{56.7}{6.8}{58.6}{100} \\
            TF-IDF/KNN & 5 & 0.500 & 0.000 & \quart{50}{0}{50}{100} \\
            TF-IDF/SVM & 5 & 0.500 & 0.000 & \quart{50}{0}{50}{100} \\
            LSTM & 6 & 0.500 & 0.000 & \quart{50}{0}{50}{100} \\
            \texttt{fastText}& 7 & 0.500 & 0.000 &  \quart{50}{0}{50}{100}\\
            TF-IDF/RF & 8 & 0.500 & 0.000 & \quart{50}{0}{50}{100} \\
            \texttt{TextCNN} & 9 & 0.500 & 0.000 & \quart{50}{0}{50}{100} \\
        \end{tabular} &
        \begin{tabular}{lcccl}
              \cellcolor[gray]{1}\textbf{Classifier} & \cellcolor[gray]{1}\textbf{Rank} &     \hphantom{\cellcolor[gray]{1}\textbf{Med}} & \hphantom{\cellcolor[gray]{1}\textbf{IQR}} & \hphantom{\cellcolor[gray]{1}}\\
            \hline
            \cellcolor{blue!10}\approach& \cellcolor{blue!10}6&\hphantom{0.000}&\hphantom{0.000}&\hphantom{\quart{50}{9.4}{50}{100}}  \\
            TF-IDF/NB & 10&\hphantom{0.000}&\hphantom{0.000}&\hphantom{\quart{50}{9.4}{50}{100}}  \\
            TF-IDF/MLP & 13&\hphantom{0.000}&\hphantom{0.000}&\hphantom{\quart{50}{9.4}{50}{100}}  \\
            \texttt{auto-fastText} & 18&\hphantom{0.000}&\hphantom{0.000}&\hphantom{\quart{50}{9.4}{50}{100}}  \\
            TF-IDF/SVM & 27&\hphantom{0.000}&\hphantom{0.000}&\hphantom{\quart{50}{9.4}{50}{100}}  \\
            TF-IDF/KNN & 30&\hphantom{0.000}&\hphantom{0.000}&\hphantom{\quart{50}{9.4}{50}{100}}  \\
            TF-IDF/RF & 30&\hphantom{0.000}&\hphantom{0.000}&\hphantom{\quart{50}{9.4}{50}{100}}  \\
            LSTM & 31&\hphantom{0.000}&\hphantom{0.000}&\hphantom{\quart{50}{9.4}{50}{100}}  \\
            \texttt{TextCNN} & 31&\hphantom{0.000}&\hphantom{0.000}&\hphantom{\quart{50}{9.4}{50}{100}}  \\
            \texttt{fastText}& 41&\hphantom{0.000}&\hphantom{0.000}&\hphantom{\quart{50}{9.4}{50}{100}}  \\
        \end{tabular}\\
      \textbf{(d). \texttt{MXNet}}  & \textbf{(e). \texttt{Caffe}} & \textbf{(f). Total rank}\\
    \end{tabular}
  \end{adjustbox}
   \end{center}
   \vspace{-0.5cm}
\end{table*}

\subsection{\approach~against State-of-the-arts}

The results for \textbf{RQ3} have been shown in Table~\ref{tab:exp3}, from which we see that, clearly, \approach~is able to considerably outperform the state-of-the-art counterparts by having a total rank of 6. In particular \approach~has been ranked as the 1st for 4 out of the 5 DL frameworks (three of them are the sole best classifier) with up to 71\% AUC improvement (e.g., compared with TF-IDF/SVM on \texttt{TensorFlow}), which is a remarkable result. The magnitudes of improvements are also significant since the median and IQR are often much better than the others overall. Further, the result with difference metrics in Table~\ref{tab:metrics} also shows that \approach~performs better than the state-of-the-art counterparts. In summary, we say:

\begin{quotebox}
   \noindent
   \textit{\textbf{\underline{To RQ3:}} \approach~is superior to the state-of-the-art counterparts with 4 out of 5 as the best, up to 71\% AUC improvement, and considerably higher recall, precision and F-measure.}
\end{quotebox}

\subsection{Why \approach~Works}

To answer \textbf{RQ4}, we look into the actual GitHub reports in the testing data for which \approach~classifies correctly but all other state-of-the-art ones made mistakes. The following \textit{Case 1} is a false positive case study for most other state-of-the-art classifiers but has been classified by \approach~correctly.

\begin{table}[!htbp]
\centering
  \begin{tabular}{lp{6cm}}
    \toprule
    \cellcolor{black!20}\textbf{Case 1:}&\cellcolor{black!20}Report \#6688 for \texttt{Keras} \\ \hline
      \textbf{Title:}&Keras accuracy is not increasing \\ \hline
   \textbf{Description:}&$\ldots$ I tried different optimizers, activation functions, number of layers, but the accuracy is reaching $\ldots$ \\ \hline
  \textbf{Comment:}&You need to take care of input numerical scale. Try to normalize every feature dimension into...\\
  \bottomrule
\end{tabular}

\end{table}

From the \textit{title} and \textit{description}, it appears that this is indeed a \bugtype bug related report. However, an immediate comment from a participant implies that the issue can be in fact due to misuse of the code (which is indeed confirmed to be true thereafter). This indicates the importance of considering the right multifaceted information that is available.

\begin{table}[t!]
\centering
  \caption{Average recall, precision and F-measure score for all classifiers over five projects. The best is in \setlength{\fboxsep}{1.5pt}\colorbox{blue!10}{blue}.}
  \label{tab:metrics}
  \footnotesize

  \begin{tabular}{l c c c c}
    \toprule
    \textbf{Classifier}&\textbf{Recall}&\textbf{Precision} & \textbf{F-measure}\\
    \midrule
    \cellcolor{blue!10}\approach &\cellcolor{blue!10}0.763 \cellcolor{blue!10}&\cellcolor{blue!10}0.804 &\cellcolor{blue!10}0.771\\
    TF-IDF/NB & 0.721 & 0.617 & 0.530\\
    TF-IDF/MLP & 0.661 & 0.745 & 0.679\\
    \texttt{auto-fastText} & 0.634 & 0.759 & 0.659\\
    TF-IDF/SVM & 0.554 & 0.551 & 0.537\\
    TF-IDF/$k$NN & 0.536 & 0.671 & 0.526\\
    TF-IDF/RF & 0.572 & 0.717 & 0.578\\
    LSTM & 0.551 & 0.559 & 0.536\\
    \texttt{TextCNN} & 0.571 & 0.616 & 0.568\\
    \texttt{fastText} & 0.500 & 0.425 & 0.459\\

  \bottomrule
\end{tabular}

\end{table}

Yet, in another case study below (\textit{Case 2}), \approach~classifies it correctly but this becomes the false negative for most other state-of-the-art classifiers.

\begin{table}[h!]
\centering
  \begin{tabular}{lp{6cm}}
    \toprule
    \cellcolor{black!20}\textbf{Case 2:}&\cellcolor{black!20}Report \#19277 for \texttt{Tensorflow} \\ \hline
      \textbf{Title:}&SSD mobilenet inference is slower w/ MKL \\ \hline
   \textbf{Description:}&$\ldots$ w/ MKL, benchmark$\_$model got 18.98B FLOPs/second, w/o MKL, it got 25.61B. From the benchmark$\_$model results, we could see that $\_$MklConv2DWithBias is the culprit$\ldots$ \\ 
  \bottomrule
\end{tabular}

\end{table}

While the \textit{title} has the word \texttt{\textcolor{blue}{`slower'}} which may be helpful, the \textit{description}, however, contains unusual terms and mixed meaning, leading to strong noises which confuse the sate-of-the-art classifiers to make a wrong prediction. In contrast, thanks to the hierarchical attentions, \approach~is able to focus on the most important part of the GitHub report and ignore the other noises, which correctly identifies it as a \bugtype bug related report.

Overall, we answer \textbf{RQ4} as:

\begin{quotebox}
   \noindent
   \textit{\textbf{\underline{To RQ4:}} The effectiveness of \approach~is the result of better semantic information handling via hierarchical attentions and the multifacetedness, which are precisely the challenges we seek to tackle in this work.}
\end{quotebox}

\section{Threats to Validity}
\label{sec:tov}

Threats to \textbf{internal validity} can be related to the parameters. To mitigate such in this work, we set the parameter of \approach~as default values or according to existing work~\cite{DBLP:conf/naacl/YangYDHSH16,DBLP:conf/nips/VaswaniSPUJGKP17}. For state-of-the-art classifiers, we tune their hyperparameters in each run whenever possible. \textbf{Construct validity} may be subject to threats due to the metrics used in the evaluation. We use AUC --- a robust metric --- to assess the performance of classifiers on 30 repeated runs in this work. To ensure statistical significance, Scott-Knott test (underpinned by hypothesis test and $\hat{A}_{12}$) is used to further verify the results. To mitigate threats to \textbf{external validity}, we conduct experiments on five popular DL frameworks with in-depth ablation analysis for \approach, including different schemes to handle the \textit{content} feature and the best combination of them for considering the multifacetedness. We have also compared \approach~with nine other state-of-the-art classifiers in the field. However, we agree that more subject and DL frameworks may prove fruitful.

\section{Related work}
\label{sec:related}

Here, we discuss the related work in light of \approach.

\textbf{Bug report identification with statistical learning.} Traditional statistical learning has been widely applied for bug report identification~\cite{DBLP:journals/alr/Polpinij21,alharthi2021efficient,DBLP:conf/icicm/OtoomAH19,kukkar2018supervised,DBLP:journals/isse/PandeySHS17,DBLP:conf/icsm/TerdchanakulHPM17,DBLP:conf/issre/WenYH16,DBLP:conf/esem/WangCWW16,DBLP:journals/infsof/XiaLSWY15}, such as identifying long-lived bug reports \cite{DBLP:journals/infsof/GomesTC21}, configuration bug reports \cite{DBLP:conf/issre/WenYH16}, security bug reports \cite{DBLP:journals/infsof/JiangLSW20,das2018security,DBLP:journals/ese/ShuXCWM21,alharthi2021efficient}, high impact bug reports \cite{DBLP:journals/infsof/WuZCZYM21,DBLP:conf/icsm/KashiwaYKO14,DBLP:journals/jcst/YangLXHS17,DBLP:conf/compsac/YangLHXS16}, or whether a report is bug-related \cite{DBLP:journals/alr/Polpinij21,DBLP:conf/icicm/OtoomAH19,kukkar2018supervised,DBLP:journals/isse/PandeySHS17,DBLP:conf/icsm/TerdchanakulHPM17,DBLP:conf/esem/WangCWW16,DBLP:journals/infsof/XiaLSWY15}. 

However, a major limitation with those works is the inability to handle semantic information, which is an important aspect to consider in bug report identification. This is one of the challenges we seek to overcome with \approach.

\textbf{Bug report identification with deep learning.} More recently, deep learning based classifiers have been proposed to tackle the limitation of semantic learning~\cite{DBLP:journals/infsof/WuZCZYM21,DBLP:conf/icsm/KallisSCP19,DBLP:journals/ese/HerboldTT20,DBLP:conf/icmla/0003FWBL18,gondaliya2018learning,DBLP:journals/tr/UmerLI20,jayagopal2021bug,DBLP:conf/internetware/QinS18,DBLP:conf/emnlp/Kim14}. 
Nevertheless, the above did not consider the fact that different parts of the reports contribute differently in identifying the report, and that multifaceted types of features exist in a GitHub report.

\textbf{HAN in software data analytic.} HAN has been deemed as an effective model for different software engineering tasks, e.g., learning the representation of code methods~\cite{wang2020reinforcement}, code summarization~\cite{DBLP:conf/pepm/XuZWCGX19}, and defect prediction~\cite{DBLP:conf/issta/ZengZZZ21}.

Our work differs from all the above as we focus on \bugtype bug related report identification for DL frameworks, with specific consideration of the challenges and characteristics of the problem contexts, which motivates the design of \approach~and MHAN therein. 
\section{Conclusion}
\label{sec:con}

In this paper, we propose \approach, an end-to-end tool for automatically identifying \bugtype bug related reports for DL frameworks. By extending HAN, \approach~is underpinned by MHAN that tackles three unresolved challenges: (1) learning the semantic knowledge in the identification and doing so by (2) taking the hierarchy of GitHub reports into account and discriminating the most important part; while (3) considering the multifacetedness of feature types. 

We verify \approach~over a dataset of 3,712 GitHub reports from five popular DL frameworks. The results reveal that:

\begin{itemize}
    \item The \textit{title} and \textit{description} of a GitHub reports should be concatenated together as part of the \textit{content} feature.
    \item The \textit{content}, \textit{comment}, and \textit{code} is the best combination for \approach~and they are considerably better than the classic HAN where only the \textit{content} is used.
    \item \approach~significantly outperforms state-of-the-art classifiers.
\end{itemize}

In future work, we plan to look into multi-label classification of the GitHub reports for DL frameworks and extend \approach~to consider more structural information of the provided code in a report, e.g., its Abstract Syntax Tree.

\footnotesize{
\bibliographystyle{IEEEtranN}
\bibliography{main}
}

\end{document}